\documentclass[aps,prd,preprint,showpacs]{revtex4}
\usepackage{graphicx}
\usepackage{subfigure}
\usepackage{hyperref}
\begin{document}
\draft
\title{Anomalous $HZ\gamma$ couplings in photon-induced collisions at the LHC}

\author{A. Senol}

\email{asenol@kastamonu.edu.tr}

\affiliation{Kastamonu University, Department of Physics, 37100,
Kuzeykent, Kastamonu, Turkey}

\author{A. T. Tasci}

\email{atasci@kastamonu.edu.tr}

\affiliation{Kastamonu University, Department of Physics, 37100,
Kuzeykent, Kastamonu, Turkey}

\author{I. T. Cakir}

\email{ilkayturkcakir@aydin.edu.tr}

\affiliation{Istanbul Aydin University, Department of Electrical and
Electronics Engineering, 34295, Sefakoy, Istanbul, Turkey}

\author{O. Cakir}

\email{ocakir@science.ankara.edu.tr}

\affiliation{Ankara University, Department of Physics, 06100,
Tando\v{g}an, Ankara, Turkey }


\begin{abstract}
We have examined $HZ\gamma$ vertex to obtain the limits on anomalous
$a_\gamma$, $b_\gamma$ and $\tilde b_\gamma$ couplings in a model
independent way through the $\gamma p$ collisions via the process
$pp\to p \gamma p\to pHqX$. The sensitivities to the anomalous
couplings can be obtained as $|b_\gamma|$, $|\tilde{b}_\gamma|\sim
10^{-3}$ for the integrated luminosity of $L_{int}=100
~\mathrm{fb}^{-1}$ at the LHC with $\sqrt{s}=14$ TeV.

\end{abstract}


\maketitle

\section{introduction}

After the discovery of Higgs boson by the ATLAS and CMS
Collaborations at the LHC \cite{Aad:2012tfa,Chatrchyan:2012ufa},
properties of Higgs boson have been studied extensively. Higgs boson
couplings to the Standard Model (SM) particles are important because
these couplings may give some hints for physics beyond the SM.
Tree-level neutral bosons coupling for $HZ\gamma$ vanish within the
electroweak interactions. Therefore, any detected signals of these
tree-level couplings would indicate the existence of new physics. In
a common scenario beyond the SM, the corrections from higher
dimensional operators can be calculated through an effective
Lagrangian. Being consistent with Lorentz and gauge invariance, the
general structure for $HZ\gamma$ vertex can be written as
\cite{Hagiwara:2000tk,Biswal:2005fh},
\begin{equation}\label{couplings} \Gamma^\gamma_{\mu\nu} = g_Z M_Z \left[
a_\gamma\,g_{\mu\nu} + \frac{b_\gamma}{M_Z^2}\,
    (q_{1\nu}q_{2\mu} - g_{\mu\nu}q_1\cdot q_2) +
    \frac {\tilde b_\gamma}{M_Z^2}\,\epsilon_{\mu\nu\alpha\beta}
    q_1^\alpha q_2^\beta\right],
\end{equation}
where $q_1$ and $q_2$ represent momentum of photon and $Z$ boson,
respectively. $M_Z$ is the mass of $Z$ boson;
$g_Z=e/\cos\theta_W\sin\theta_W$; $a_\gamma$, $b_\gamma$ and $\tilde
b_\gamma$ are constants which can be written in form factors. Here,
$a_\gamma$ vanishes according to the electromagnetic gauge
invariance for $q_1^2=0$.

In the photon-induced processes a quasi-real photon emitted from one
of the proton beam \cite{Budnev:1974de,Ginzburg:1981vm} can be
described in the framework of equivalent photon approximation (EPA).
For a process in a photon induced $\gamma p$ collision can be
different from pure deep inelastic scattering process by means of
two experimental signatures emerge in the following way
\cite{Piotrzkowski:2000rx,Royon:2007ah,Albrow:2008pn}: First,
forward detectors can detect the particles with large
pseudorapidity. If the proton emits a photon, it scatters with a
large pseudorapidity and can not be detected from the central
detectors. However, potential forward detectors located at 220 m and
420 m away from the interaction point can detect the particles with
large pseudorapidity providing some information on the scattered
proton energy. Second, if the photon emitting intact protons exits
the central detector without being detected, the energy deposit in
the forward region decreases compared to the case in which the
proton remnant is detected by the colorimeters. Accordingly, one of
the forward regions of the central detector has a large energy
deficiency. The region devoid of particles defines forward rapidity
gaps. $pp$ backgrounds from deep inelastic processes can be
seperated by applying a selection cut on this quantity.

Photon induced processes at the LHC have been studied as a probe of
new physics beyond the SM
\cite{Atag:2000tn,Kepka:2008yx,deFavereaudeJeneret:2009db,Chapon:2009hh,Albrow:2010yb,Sahin:2011yv,Sahin:2012mz,Sahin:2012ry,Senol:2013ym}.
Extensive studies on determining the sensitivity to the $HZ\gamma$
vertex have been performed at $e\gamma$
\cite{Gabrielli:1997ix,Cotti:1997er}, $e^-e^+$
\cite{Hagiwara:2000tk,Dutta:2008bh,Rindani:2009pb,Rindani:2010pi,GutierrezRodriguez:2010ri}
and $pp$ \cite{Hankele:2006ma} collisions.

In this work, we have studied anomalous $HZ\gamma$ vertex to obtain
the limits on $a_\gamma$, $b_\gamma$ and $\tilde b_\gamma$ couplings
in a model independent way through a $\gamma p$ collision via the
process $pp \to p\gamma p \to pHqX$ at the LHC. For all
calculations, we use the computer package CalcHEP
\cite{Belyaev:2012qa} by implementing the anomalous interaction
vertices given in Eq. \ref{couplings}.

We consider the acceptance regions of the ATLAS and CMS forward
detectors to tag the protons with some energy loss fraction
$\xi=E_{loss}/E_{beam}$. ATLAS and CMS experiments will have forward
detectors with the acceptance of $0.0015<\xi<0.15$
\cite{Royon:2007ah,Albrow:2008pn} within the pseudorapidity ranges
$9.5<\eta<13$. When the forward detectors are installed closer to
the interaction points, higher $\xi$ value is obtained. Therefore,
we define the parameter sets corresponding to these acceptances and
pseudorapidity ranges. The parameters can also be related to the
transverse momentum cuts on the tag protons as,
\begin{equation}\label{pt}
p_T=\frac{\sqrt{E_p^2(1-\xi)^2-m_p^2}}{\cosh{\eta}}
\end{equation}
where $E_p$ and $m_p$ are the energy and mass of proton,
respectively. We define the parameter sets, namely PI, PII and PIII
within the pseudorapidity range $9.5<\eta<13$ for minimal transverse
momentum cuts on the scattered protons with $p_T>0.03$ GeV,
$p_T>0.1$ GeV and $p_T>0.2$ GeV, respectively.

\section{Decay width for $HZ\gamma$}

The decay width of $H\to Z\gamma$ with anomalous couplings can be
calculated considering the effective vertex in Eq. \ref{couplings}
as,
\begin{eqnarray}
\Gamma(H\to Z\gamma)=\frac{\alpha M_Z^2(M_H^2-M_Z^2)}{8 c_W^2
s_W^2M_H^3M_Z^4}\left[8a_\gamma^2M_Z^4+6a_\gamma b_\gamma
M_Z^2(M_Z^2-M_H^2)+(b_\gamma^2+\tilde{b}_\gamma^2)(M_H^2-M_Z^2)^2\right]
\label{eq:decay}
\end{eqnarray}
where $s_W\equiv \sin\theta_W$ and $c_W\equiv \cos\theta_W$, $M_H$
denotes mass of the Higgs boson. There is also contribution from one
loop diagram including $HZ\gamma$ anomalous couplings. However, this
contribution is proportional to the power of six of the anomalous
couplings. Taking into account the experimental bounds on these
couplings, which are smaller than $10^{-2}$, tree level contribution
dominates.
 \begin{figure}
\centering
\includegraphics[width=10cm]{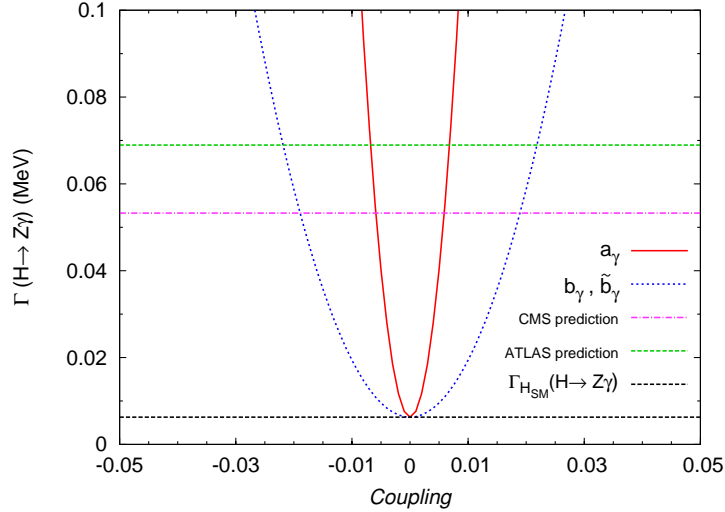}
\caption{Decay width for $H\to Z\gamma$ depending on the anomalous
couplings $a_\gamma$, $b_\gamma$, $\tilde{b}_\gamma$.}\label{decay}
\end{figure}

The decay width of $H\to Z\gamma$ depending on the anomalous
couplings is given in Fig.~\ref{decay}. In Fig.~\ref{decay}, we set
two of these anomalous couplings equal to zero while varying the
other coupling changes in the range $[-0.05,0.05]$. The horizontal
dashed line (black) denotes the one loop calculation of the decay
width $H\to Z\gamma$, which is $6.27\times 10^{-3}$ MeV
\cite{Denner:2011mq} for the Higgs boson mass of 125 GeV within the
framework of the SM. The limits on the anomalous couplings can be
extracted from the intersection points of the experimental
predictions and the calculated values of decay width $H \to
Z\gamma$. It is seen from Fig. \ref{decay} that the prediction from
ATLAS \cite{Aad:2012tfa} experiment limits these anomalous couplings
as $|a_\gamma|<0.007$ and $|b_\gamma|$, $|\tilde{b}_\gamma|<0.022$.
The CMS \cite{Chatrchyan:2012ufa} prediction leads to more strict
bounds. Due to more restriction on anomalous coupling $a_\gamma$ we
assume $a_\gamma=0$, while other couplings allowed to changed in the
considered range.

\section{cross section for single higgs production}

\begin{figure}
\centering
\includegraphics[width=14cm]{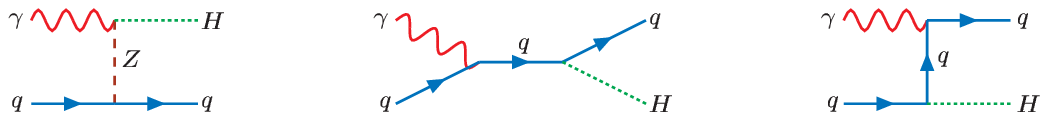}
\caption{Feynman diagrams for the subprocess $\gamma q \to
Hq$.}\label{fig1}
\end{figure}

The contributing tree level Feynman diagrams for the single
production of Higgs boson through $\gamma q \to Hq$ subprocess
(where $q=u, \bar{u}, d, \bar{d}, s, \bar{s}, c, \bar{c}, b,
\bar{b}$) are shown in Fig.~\ref{fig1}. The differential cross
section for the subprocess $\gamma q \to Hq$ is obtained as
neglecting quark masses,
\begin{eqnarray}
\frac{d\hat{\sigma}}{d\hat{t}}&=&\frac{\pi^2\alpha^2(12s_W^2-8s_W^4-9)\hat{t}}
{18c_W^4s_W^4M_Z^2[(\hat{t}-M_Z^2)^2+\Gamma_Z^2M_Z^2]}[4a_\gamma^2M_Z^4+
4a_\gamma b_\gamma M_Z^2(\hat{t}-M_H^2)\\\nonumber&&+(b_\gamma^2
+\tilde{b}_\gamma^2)[M_H^4+\hat{t}^2+2(\hat{s}-2M_H^2)(\hat{s}
+\hat{t})]].\label{eq:2}
\end{eqnarray}
We find the total cross section of $pp \to p\gamma p \to pHqX$
process by integrating differential cross section of $\gamma q \to
Hq$ subprocess over the parton distribution functions CTEQ6L
\cite{Pumplin:2002vw} and photon spectrum in EPA
\cite{Budnev:1974de,Ginzburg:1981vm,Piotrzkowski:2000rx}, by using
the CalcHEP.

\begin{figure}
\centering
\includegraphics[width=10cm]{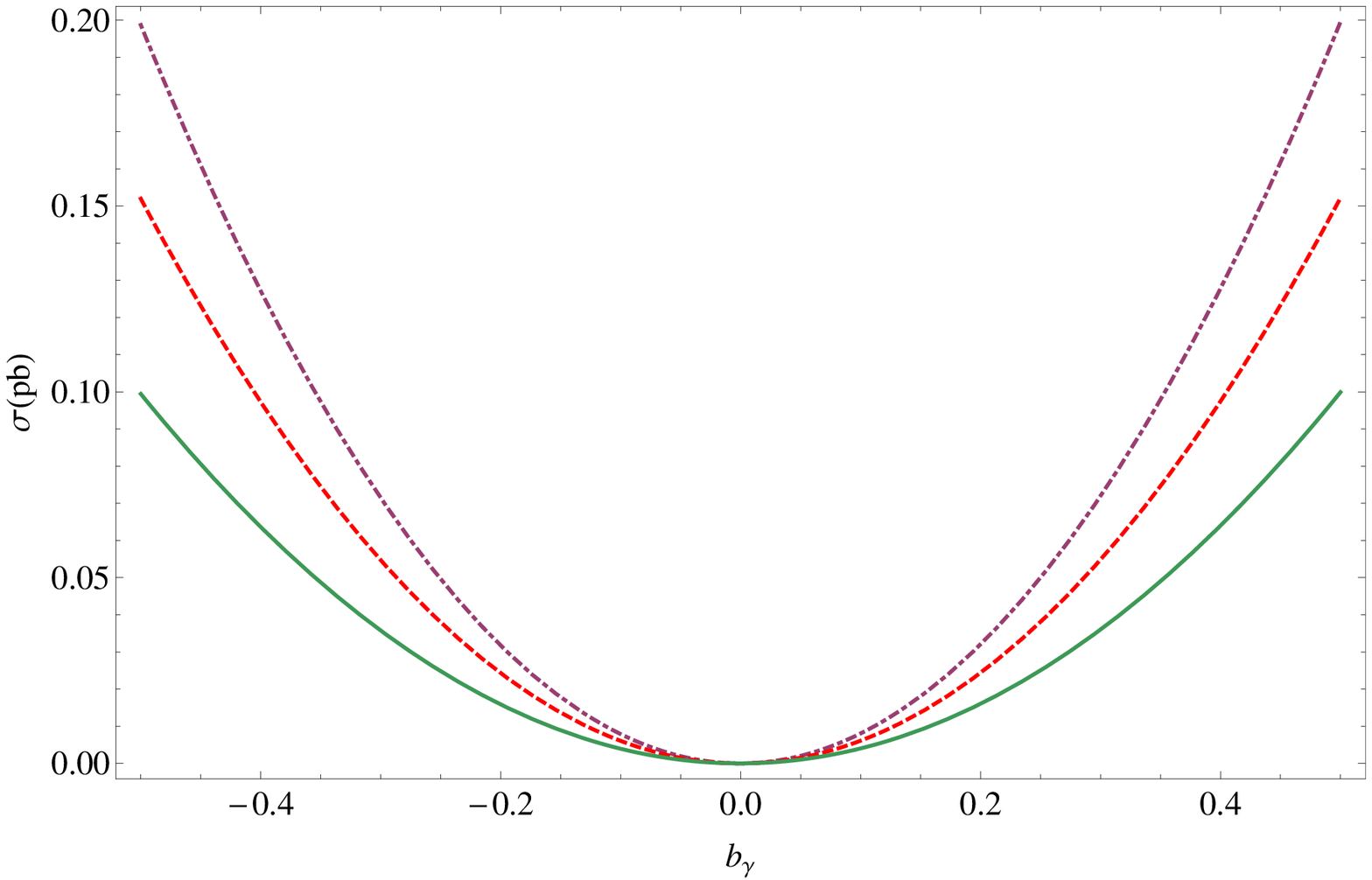} \caption{The
total cross section depending on anomalous coupling $b_\gamma$ with
$\sqrt{s}=14$ TeV. Dot-dashed, dashed and solid lines correspond to
parameter sets PI, PII and PIII, respectively.}\label{fig:csb}
\end{figure}

\begin{figure}
\centering
\includegraphics[width=10cm]{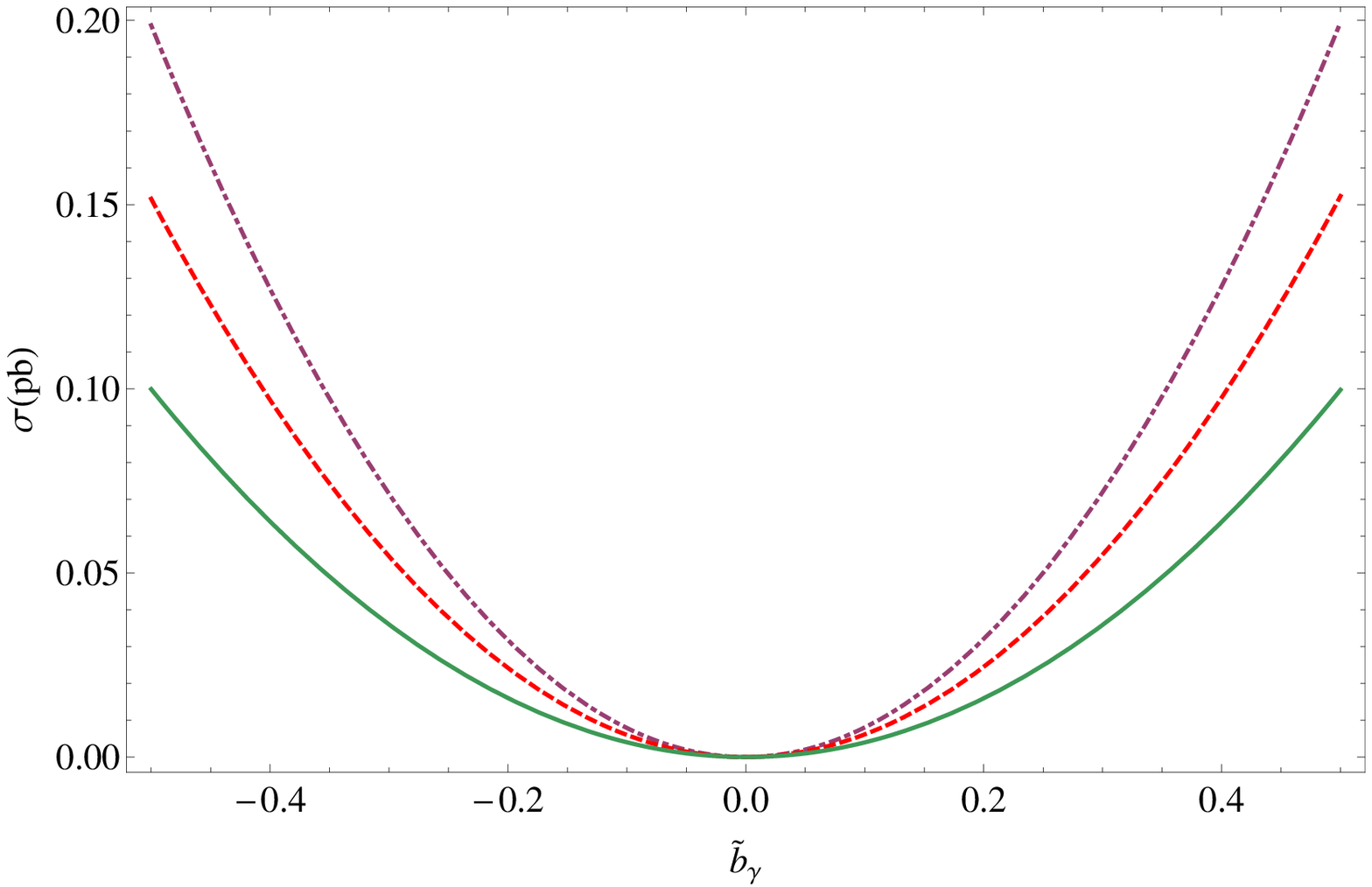} \caption{The
total cross section depending on anomalous coupling
$\tilde{b}_\gamma$ with $\sqrt{s}=14$ TeV. Dot-dashed, dashed and
solid lines correspond to parameter sets PI, PII and PIII,
respectively.}\label{fig:csbt}
\end{figure}

In Figs.~\ref{fig:csb} and \ref{fig:csbt} the total cross sections
of $pp \to p\gamma p \to pHqX$ process as a function of $b_\gamma$
and $\tilde{b}_\gamma$ are illustrated for the parameter sets PI,
PII, PIII.

\section{sens\i{}t\i{}v\i{}ty to anomalous coupl\i{}ngs}

The $\chi^2$ test for the bounds of anomalous $b_{\gamma}$ and
$\tilde{b}_{\gamma}$ couplings at 95 \% C.L. can be written as
\begin{eqnarray}
\chi^{2}=\left(\frac{\sigma_{SM}-\sigma_{AN}}{\sigma_{SM} \,\,
\delta}\right)^{2}
\end{eqnarray}
where $\sigma_{AN}$ is the cross section including considered
anomalous couplings; $\delta=\frac{1}{\sqrt{N}}$ is the statistical
error and here $N$ is the number of SM events. The number of events
are given by $N=\sigma_{SM}\times L_{int}\times BR(H\to b\bar
b)\times (\epsilon_{b-tag})^2$ where $\epsilon_{b-tag}$ denotes the
$b$-tagging efficiency and $L_{int}$ is the integrated luminosity.
We use kinematical cuts for transverse momentum of final state
quarks to be $p_T^{j}> 15$ GeV and pseudorapidity to be $|\eta^{j}|<
2.5$.

In Tables \ref{tab:tab1}, \ref{tab:tab2} and \ref{tab:tab3}, we have
listed 95\% C.L. sensitivity limits on the couplings $b_{\gamma}$
and $\tilde{b}_{\gamma}$ for various integrated luminosities by
varying one coupling at a time, for sets of minimal transverse
momentum cuts PI, PII and PIII, respectively.

\begin{table}[t]
\caption{Sensitivity (95\% C.L.) to anomalous $HZ\gamma$ couplings
for parameter set PI in the $\gamma p$ collisions at the LHC with
$\sqrt{s}=14$ TeV for various integrated luminosities.
\label{tab:tab1}}
\begin{tabular}{c@{\hspace{0.3cm}}c@{\hspace{0.3cm}}c@{\hspace{0.3cm}}c}
  \hline\hline
  L($\mathrm{fb}^{-1}$) & $b_\gamma$ & $\tilde{b}_\gamma$ \\\hline
  50  &  ($-9.93\times 10^{-3}$, $8.89\times 10^{-3}$) & ($-9.92\times 10^{-3}$, $8.90\times 10^{-3}$) \\
  100 &  ($-8.44\times 10^{-3}$, $7.39\times 10^{-3}$) & ($-8.42\times 10^{-3}$, $7.41\times 10^{-3}$) \\
  200 &  ($-7.19\times 10^{-3}$, $6.14\times 10^{-3}$) & ($-7.17\times 10^{-3}$, $6.16\times 10^{-3}$) \\
  \hline\hline
\end{tabular}
\end{table}

\begin{table}[t]
\caption{Same as Table \ref{tab:tab1}, but for parameter set PII.
\label{tab:tab2}}
\begin{tabular}{c@{\hspace{0.3cm}}c@{\hspace{0.3cm}}c@{\hspace{0.3cm}}c}
  \hline\hline
  L($\mathrm{fb}^{-1}$) &  $b_\gamma$ & $\tilde{b}_\gamma$ \\\hline
  50   & ($-9.89\times 10^{-3}$, $9.51\times 10^{-3}$) & ($-1.03\times 10^{-2}$, $9.11\times 10^{-3}$) \\
  100  & ($-8.35\times 10^{-3}$, $7.96\times 10^{-3}$) & ($-8.79\times 10^{-3}$, $7.57\times 10^{-3}$) \\
  200  & ($-7.05\times 10^{-3}$, $6.67\times 10^{-3}$) & ($-7.50\times 10^{-3}$, $6.27\times 10^{-3}$) \\
  \hline\hline
\end{tabular}
\end{table}

\begin{table}[t]
\caption{Same as Table \ref{tab:tab1}, but for parameter set PIII.
\label{tab:tab3}}
\begin{tabular}{c@{\hspace{0.3cm}}c@{\hspace{0.3cm}}c@{\hspace{0.3cm}}c}
  \hline\hline
  L($\mathrm{fb}^{-1}$) &  $b_\gamma$ & $\tilde{b}_\gamma$ \\\hline
  50   & ($-1.11\times 10^{-2}$, $9.96\times 10^{-3}$) & ($-1.03\times 10^{-2}$, $1.07\times 10^{-2}$) \\
  100  & ($-9.43\times 10^{-3}$, $8.29\times 10^{-3}$) & ($-8.61\times 10^{-3}$, $9.07\times 10^{-3}$) \\
  200  & ($-8.03\times 10^{-3}$, $6.88\times 10^{-3}$) & ($-7.20\times 10^{-3}$, $7.67\times 10^{-3}$) \\
  \hline\hline
\end{tabular}
\end{table}

\begin{figure}
\centering
\includegraphics[width=10cm]{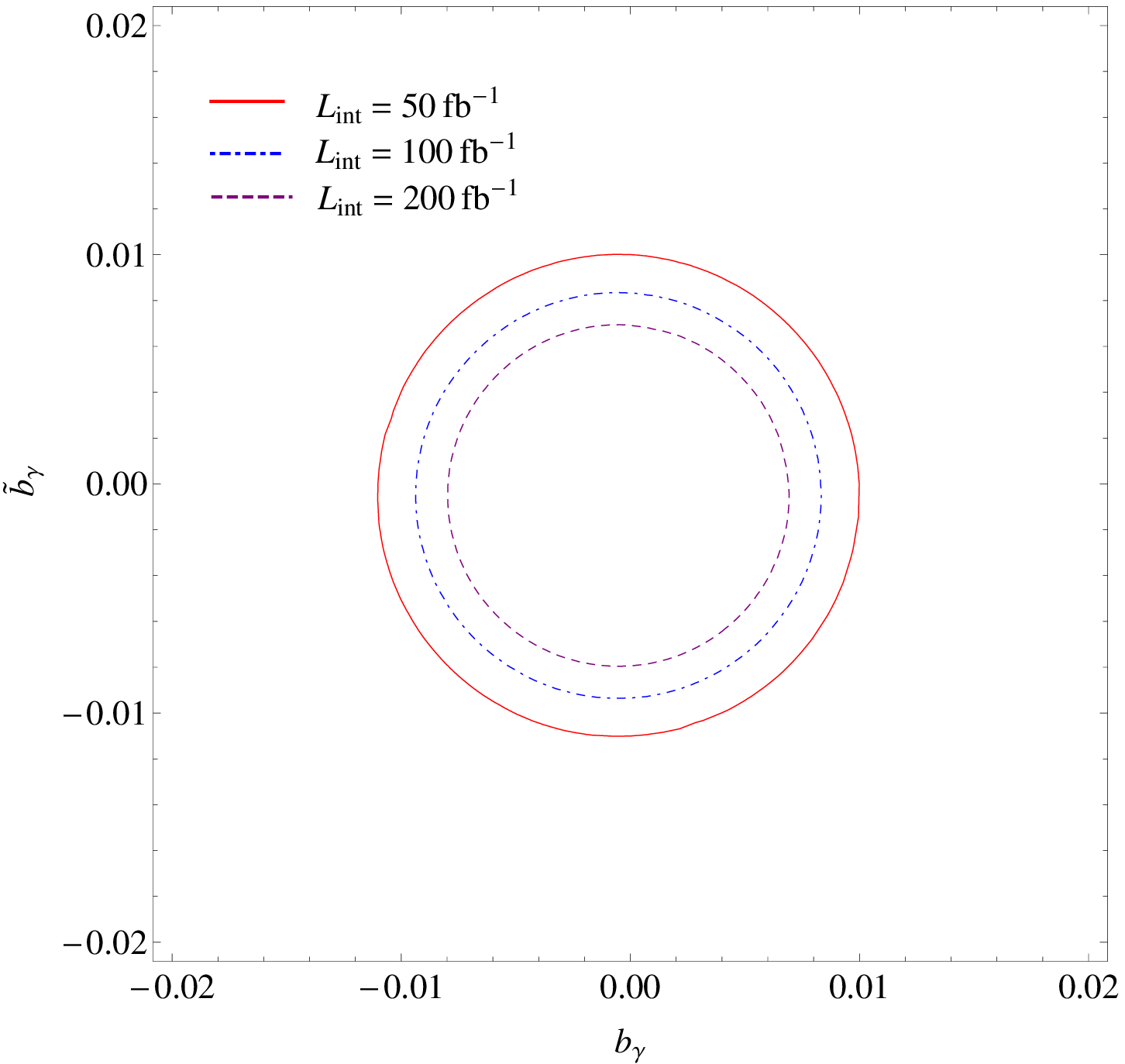} \caption{Two dimensional contour
plot for anomalous couplings $b_\gamma$ and $\tilde{b}_\gamma$ for
the process $\gamma q\to Hq$ with $\sqrt{s}=14$ TeV for parameter
set PI.}\label{fig:cont1}
\end{figure}

\begin{figure}
\centering
\includegraphics[width=10cm]{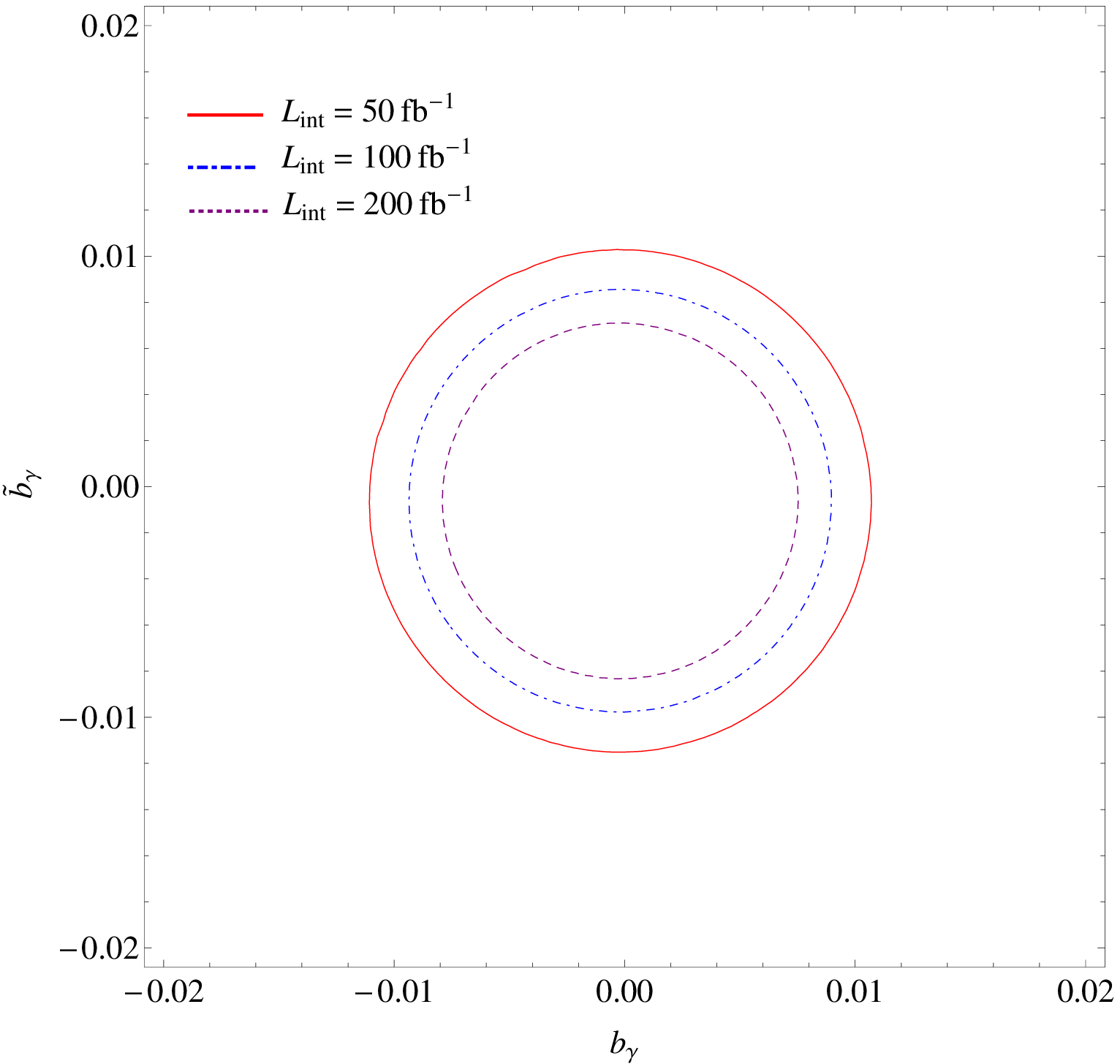} \caption{Two dimensional contour
plot for anomalous couplings $b_\gamma$ and $\tilde{b}_\gamma$ for
the process $\gamma q\to Hq$ with $\sqrt{s}=14$ TeV for parameter
set PII.}\label{fig:cont2}
\end{figure}

\begin{figure}
\centering
\includegraphics[width=10cm]{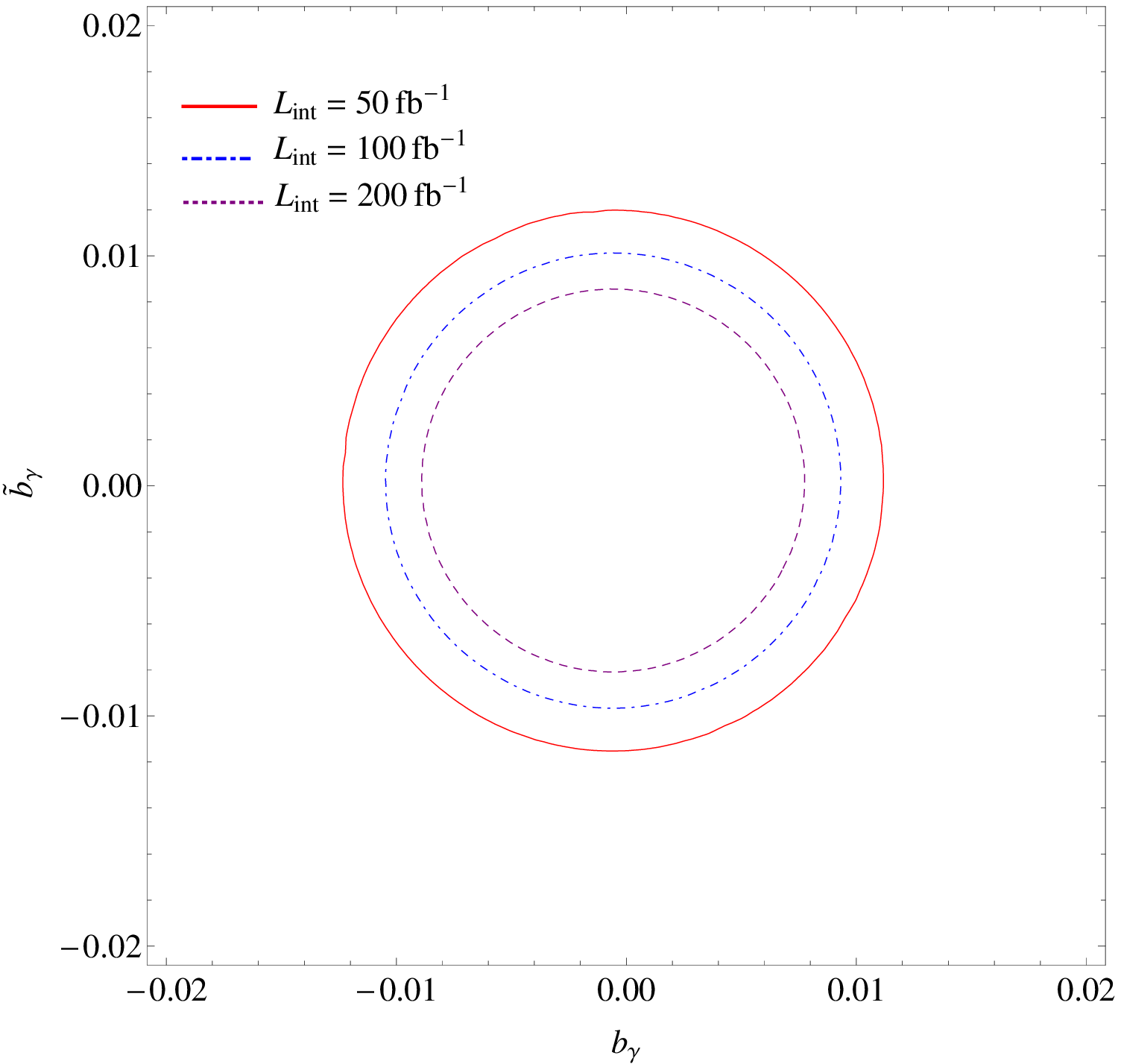} \caption{Two dimensional contour
plot for anomalous couplings $b_\gamma$ and $\tilde{b}_\gamma$ for
the process $\gamma q\to Hq$ with $\sqrt{s}=14$ TeV for parameter
set PIII.}\label{fig:cont3}
\end{figure}

The sensitivities to the anomalous couplings in $b_\gamma
-\tilde{b}_\gamma$ plane are plotted when two of the anomalous
parameters are changed independently in Figs. \ref{fig:cont1},
\ref{fig:cont2} and \ref{fig:cont3} for the parameter sets PI, PII
and PIII, respectively.
\section{Conclusions}
We have analyzed the single production of Higgs boson including
anomalous $HZ\gamma$ vertices at the photon-induced $\gamma p$
collisions at the LHC. This process gives opportunity to investigate
anomalous $HZ\gamma$ couplings $|b_\gamma|$ and $|\tilde{b}_\gamma|$
at the order of $10^{-3}$. We find the best limits on these
anomalous couplings which can be compared to the bounds $O(10^{-2})$
obtained from the associate production of $HZ$ \cite{Rindani:2010pi}
at linear collider with $\sqrt{s}=500$ GeV.

\begin{acknowledgments} A.S. would like to thank Abant Izzet Baysal
University Department of Physics where of part this study was
carried out for their hospitality.
\end{acknowledgments}

\end{document}